\documentclass[conference,letterpaper]{IEEEtran}

\usepackage{cite}
\usepackage{textcomp}
\usepackage{amsmath, amssymb, amsfonts}
\usepackage{dsfont}
\usepackage{accents}
\usepackage{algorithm}
\usepackage{color}
\usepackage{graphicx}
\usepackage{booktabs}
\usepackage{caption}
\usepackage{subcaption}
\usepackage{enumerate}
\usepackage{mathrsfs}
\usepackage{mathtools}

\usepackage{pgfplots}
\usepackage{algpseudocode}
\newtheorem{theorem}{\bf{Theorem}}
\newtheorem{remark}{\bf{Remark}}

\DeclareMathOperator*{\argmax}{arg\,max}

\DeclareMathOperator{\maximize}{maximize \ }
\DeclareMathOperator{\minimize}{minimize \ }
\DeclareMathOperator{\subjectto}{subject \ to \ }

\begin{document}
\title{Remote Estimation of Markov Processes\\over Costly Channels: On the Benefits of\\Implicit Information} 
\author{
Edoardo D. Santi, Touraj Soleymani and Deniz Gündüz\\
Department of Electrical and Electronic Engineering, Imperial College London, United Kingdom\\
{\tt\small \{edoardo.santi17, d.gunduz, touraj\} @imperial.ac.uk}
%\thanks{Edoardo D. Santi, Deniz Gunduz, and Touraj Soleymani are all with the Department of Electrical and Electronic Engineering, Imperial College London, London SW7 2AZ, United Kingdom (e-mails: {\tt\small eds17@ic.ac.uk}, {\tt\small d.gunduz@imperial.ac.uk}, {\tt\small touraj@imperial.ac.uk}).}%
}
    % \IEEEauthorblockN{Edoardo D. Santi, Deniz Gunduz, and Touraj Soleymani}
    % \IEEEauthorblockA{Department of Electrical and Electronic Engineering\\
    %                    Imperial College London\\
    %                    London, UK\\
    %                    Email: {\tt\small eds17@ic.ac.uk}, {\tt\small d.gunduz@imperial.ac.uk}, {\tt\small touraj@imperial.ac.uk} }} 
  %\and

%%% Many authors with many affiliations:
% \author{%
%   \IEEEauthorblockN{Andrew R.~Barron\IEEEauthorrefmark{1},
%                     Claude E.~Shannon\IEEEauthorrefmark{2},
%                     David Slepian\IEEEauthorrefmark{2},
%                     and Jacob Ziv\IEEEauthorrefmark{2}\IEEEauthorrefmark{3}}
%   \IEEEauthorblockA{\IEEEauthorrefmark{1}%
%                    Department of Statistics and Data Science, Yale University, New Haven, CT, USA,
%                     andrew.barron@yale.edu}
%   \IEEEauthorblockA{\IEEEauthorrefmark{2}%
%                     Bell Telephone Laboratories, Inc.,
%                     Murray Hill, NJ, USA,
%                     \{csh,dsl,jz\}@bell-labs.com}
%   \IEEEauthorblockA{\IEEEauthorrefmark{3}%
%                     Department of Electrical Engineering, Technion---Institute of Technology, Haifa, Israel,
%                     jz@ee.technion.ac.il}
% }

\maketitle

%%%%%%
%% Abstract: 
%% If your paper is eligible for the student paper award, please add
%% the comment "THIS PAPER IS ELIGIBLE FOR THE STUDENT PAPER
%% AWARD." as a first line in the abstract. 
%% For the final version of the accepted paper, please do not forget
%% to remove this comment!
%%

\begin{abstract}
In this paper, we study the remote estimation problem of a Markov process over a channel with a cost. We formulate this problem as an infinite horizon optimization problem with two players, i.e., a sensor and a monitor, that have distinct information, and with a reward function that takes into account both the communication cost and the estimation quality. We show that the main challenge in solving this problem is associated with the consideration of implicit information, i.e., information that the monitor can obtain about the source when the sensor is silent. Our main objective is to develop a framework for finding solutions to this problem without neglecting implicit information \textit{a priori}. To that end, we propose three different algorithms. The first one is an alternating policy algorithm that converges to a Nash equilibrium. The second one is an occupancy-state algorithm that is guaranteed to find a globally optimal solution. The last one is a heuristic algorithm that is able to find a near-optimal solution.
\end{abstract}
% \begin{IEEEkeywords}
% cyber-physical systems, globally optimal solutions, implicit information, Nash equilibria, Markov processes, optimal policies, remote estimation.
% \end{IEEEkeywords}

\section{Introduction}
Cyber-physical systems integrate physical and digital components into unified intelligent systems \cite{kim_cyberphysical_2012}. These  systems, which capitalize on computation, communication, and control to enhance the overall performance, have various applications in smart cities, smart factories, smart healthcare, and smart transportation. However, it is essential to acknowledge the dynamic nature of cyber-physical systems, which necessitates persistent monitoring to capture real-time information about the physical environment. This constant stream of real-time data empowers cyber-physical systems to swiftly adapt to changing conditions, ensuring that decisions are made based on the most pertinent and up-to-date information~\cite{soleymani_value_2022, soleymani_value_2023, touraj-thesis, uysal2022semantic, gunduz2022semantic}.

In this paper, we study the remote estimation problem of Markov processes over costly channels. In particular, we consider a setting where there is a networked system composed of a sensor observing a Markov source and a remote monitor that needs to be informed about the state of the source. The sensor should transmit observed information to the monitor over a costly channel, and the monitor should estimate the state of the source in real time. As we will see, this problem can be quite challenging when both the communication cost and the estimation quality are taken into account. Nevertheless, we aim at developing a framework that enables us to solve it either exactly or approximately.

\subsection{Related Work}
The problem of remote estimation of Markov processes with continuous states over costly channels has been addressed in~\cite{soleymani_value_2022, soleymani_value_2023, touraj-power, erasure2023, molin2017, lipsa_remote_2011}. These works characterize the optimal solutions rigorously and shed light on the role of implicit information for Gauss-Markov processes. The remote estimation of Markov processes with discrete states over costly channels, the problem of interest in the present paper, has been addressed in \cite{chakravorty_optimal_2014, chakravorty_distortion-transmission_2015, chakravorty_fundamental_2016, pappas_goal-oriented_2021, salimnejad_state-aware_2023, salimnejad_real-time_2023, krale_act-then-measure_2023-2, bellinger_active_nodate, nam_reinforcement_2021}. Notably, closed-form threshold polices were found in \cite{chakravorty_optimal_2014} when the source has a symmetric Toeplitz transition matrix. Then, it was shown in \cite{chakravorty_distortion-transmission_2015} that a piece-wise linear convex decreasing function can represent the trade-off curve between the estimation error and the transmission rate. Under similar conditions, general properties of the optimal solution were discussed in \cite{chakravorty_fundamental_2016}. Channel noise in the context of remote actuation is studied in \cite{pappas_goal-oriented_2021}, where different types of heuristic policies are compared. Channel noise is also considered in \cite{salimnejad_state-aware_2023}, where policies for a two-state Markov chain 
are proposed, taking into account the importance of each of the two states in terms of the actions to be taken by the monitor. This work is extended in \cite{salimnejad_real-time_2023} to $N$-state Markov chains, where an optimization-based method is proposed to find the optimal parameter for a randomized stationary policy. Other works such as ~\cite{krale_act-then-measure_2023-2, bellinger_active_nodate, nam_reinforcement_2021} propose solutions for solving variations of Markov decision processes (MDPs) in which the monitor needs to pay a fixed price to either observe the current or the next state. This framework can be used when it is the monitor that requests updates. When it is the transmitting sensor that schedules transmissions, we can expect the performance to be better, as the sensor has knowledge of the states of the source. These sets of problems fall within the category of effective/pragmatic communications as defined in \cite{gunduz2022semantic, Tung:JSAC:21, gunduz2023timely}, where the state/context of the receiver becomes relevant when deciding the communication policy

\subsection{Contributions and Paper Structure}
In this paper, we tackle the remote estimation problem of sources that are discrete-state Markovian over channels that are noiseless but costly. We formulate this problem as an infinite horizon optimization problem with two players, i.e., a sensor and a monitor, that have distinct information and with a reward function that takes into account both the communication cost and the estimation quality. We show that the main challenge in solving this problem is associated with the consideration of implicit information. Our main objective is to develop a framework for finding solutions to this problem without neglecting implicit information apriori. To that end, we propose three different algorithms. The first one is an alternating policies algorithm that converges to a Nash equilibrium. The second one is an occupancy-state algorithm that is guaranteed to find a globally optimal solution. The last one is a heuristic algorithm that finds a near-optimal solution that we prove to be optimal in the special case of requiring perfect reconstruction.

The rest of this paper is structured as follows. We formulate the problem of interest mathematically in Section II. We provide the alternating policies algorithm in Section III, the occupancy-state algorithm in Section IV, and the heuristic algorithm in Section V. We then present our numerical results in Section VI. Finally, we conclude the paper in Section VII.

%Consider a discrete-time Markov chain with a finite state space $S$ and a transition probability matrix $P$, where the element $P(s,s')$ is the probability of transitioning to state $s'$ after being in state $s$. 
%The sensor should transmit observed information to the monitor over a costly channel, and the monitor should estimate the state of the Markov chain in real time. 

\section{Problem formulation}
Consider a networked system composed of a sensor observing states of a source and a remote monitor that needs to be informed about the states of the source. The source is modeled by a discrete-time finite-state Markov chain. At each time step, the sensor observes the current state of the source, and decides whether to transmit the state value to the monitor, which incurs a fixed transmission cost $c_t=1$ at time $t$ and guarantees the correct instantaneous estimation of the state at the monitor; or not to transmit, which incurs a zero cost $c_t=0$ but leaves the monitor to guess the state of the source. Consequently, a unit common reward $r_t=1$ is gained at time $t$ if the monitor's guess matches the actual state of the source, otherwise zero reward $r_t=0$ is gained. The combined reward at time $t$ is therefore given by~$R_t=r_t-\lambda \cdot c_t$, where $\lambda$ is chosen depending on which point of the trade-off curve we would like to operate on.

This remote estimation problem can be formulated formally as a two-player team game denoted by $\hat{M}=(\mathcal{I},\mathcal{S},\mathcal{A},P,Z,O,R)$, where $I=\{i_1,i_2\}$ is the set of players, i.e., the sensor and the monitor, where the latter receives observations and acts after the former has already acted; $\mathcal{S}$ is the state space of the Markov chain; $\mathcal{A}=\mathcal{A}^1 \times \mathcal{A}^2$ is the joint action space, where the action of the sensor is $a^1 \in \mathcal{A}^1 = \{0,1\}$ such that $a^1 = 1$ means to transmit and $a^1 = 0$ means not to transmit, and the action of the monitor is $a^2 \in \mathcal{A}^2 = \mathcal{S}$, which represents the state estimated by the monitor; $P$ is the transition probability matrix such that the element $P(s,s')$ represents the probability of transitioning from state $s$ to state $s'$; $Z=Z^1 \times Z^2$ is the joint observation space, where $z^1 \in \mathcal{S} = Z^1$ and $z^2 \in \mathcal{S} \cup \{ \epsilon \} = Z^2$ and $\epsilon$ is the empty observation that occurs when no message is sent; $O=\{O^1,O^2 \}$ represents the set of the observation functions, where $O^1(z^1_t,s_t)= Pr(z^1_t|s_t)=\mathds{1}[z^1_t=s_t]$ and $O^2(z^2_t,s_t,a^1_t)=Pr(z^2_t|s_t,a^1_t)=a^1_t\mathds{1}[z^2_t=s_t]+(1-a^1_t)\mathds{1}[z^2_t=\epsilon]$; and finally $R$ is the combined reward, which is equal to $1-\lambda$ when the state is estimated correctly due to a transmission, $1$ when the state is estimated correctly without a transmission, $0$ for an incorrect estimation of the state without transmission, and $-\lambda$ for an incorrect estimation despite a transmission. Note that the latter is not possible for any reasonable policy. Given a variable, we use the notation $x_{t_1:t_2}$ to denote the sequence of values that the variable takes between and including the time-steps $t_1$ and $t_2$.

We are interested in finding the transmission and estimation policies that jointly maximize the infinite horizon average reward function, i.e., we would like to solve the following optimization problem:
\begin{align}
&\hspace{-1.25cm}\text{Problem 1:} \nonumber\\
&\underset{\pi \in \Pi}{\maximize} \mathds{E}_{\sim b_0, P, \pi} \left[ \lim_{T \to \infty} \frac{1}{T} \sum_{t=0}^{T-1} R_t \right],
\end{align}
where $b_0$ is the initial distribution of the source's states, and $\Pi$ is the set of joint history-dependent stochastic policies for players 1 and 2. Note that $\Pi$ represents the most comprehensive set of achievable policies. The optimal value of Problem 1 is denoted by $J^*$.

%We aim to find policies that achieve the optimal value of $J^*$.

\section{Role of Implicit Information}
We refer to the information that the monitor can obtain about the state of the source when the sensor is silent as \textit{implicit information}. This information is relevant as the two players are jointly optimized and they are aware of each others' policies. However, jointly optimizing the players in this context is not trivial, as the optimal policy of the sensor depends on that of the monitor, and vice versa. Note that Problem 1 can be simplified if we neglect implicit information. This accordingly leads to a decoupling in the design of the sensor and the monitor. However, this approach, which neglects implicit information \textit{a priori} leads to a suboptimal solution in general as it does not take full advantage of the available information. 

In this study, we aim to devise methodologies that can find exact or approximate solutions to this problem without neglecting implicit information. More specifically, we propose three different algorithms to solve Problem 1. The first one deals with the interdependency of the transmission and estimation policies by optimizing the policy of one player while fixing the other, and then repeats this process until convergence. The second algorithm aims to achieve global optimality by recasting the original two-player problem into a single-player occupancy MDP. Finally, the third algorithm, although heuristic, is able to find an optimal solution whenever perfect reconstruction at the monitor is required.

\section{Alternating Policy Optimization}

In this section, we propose an algorithm that finds a Nash equilibrium to Problem 1. It is clear that for the sensor, the state $x=(s,s_m,n) \in \mathcal{X} = \mathcal{S}^2 \times \mathds{Z^+_0}$ is a sufficient statistic for the purpose of finding an optimal policy, where $s$ is the current state of the source, $s_m$ is the last transmitted state and $n=t-\tau$ where $t$ is the current time-step and $\tau$ is the time of the last transmission. Similarly, we set the monitor states as $y=(s_m,n) \in \mathcal{Y} = \mathcal{S} \times \mathds{Z^+_0}$, which is equivalent to the state representation of the sensor without including the current state, as it is not known by the monitor. 

Algorithm~\ref{alg:alternating} summarizes the alternating policies algorithm. Note that step $k$ of the algorithm indicates a combined optimization of both players. We fix the estimation policy $\pi^2: \mathcal{Y} \mapsto \mathcal{A}^2$ and initialize it as $(P^n)^Te_{s_m}$, which is the basic monitor policy that disregards implicit information. The Markov chain and the monitor form an MDP denoted by $M_1^k=(\mathcal{X},\mathcal{A}^1 ,R_1^k,P_1)$. The state space $\mathcal{X}$ is formed by states $x= (s, s_m, n) \in \mathcal{S}^2 \times \{ 1,2, \ldots, n_{max} \}$. The action space $\mathcal{A}^1$ is still $\{ 0,1 \}$. The reward function depends on the policy of the monitor and is defined as $R_1^k(x,0)=\mathds{1}(\pi_2^{k-1}(s_m,n)=s)$ and $R_1^k(x,1)=1-\lambda$, corresponding to the two possible actions. The transition operator is $P_1 \in \{ P_1^0, P_1^1 \}$, where $P_1^0(x,x')=P(s,s')\mathds{1}(s_m'=s_m)\mathds{1}(n'=n+1)$ and $P_1^1(x,x')=P(s,s')\mathds{1}(s_m'=s)\mathds{1}(n'=0)$, corresponding to the two possible actions. We constrain $n$ to never exceed a maximum value $n_{max}$. To do so, we modify the reward and transition functions when $n=n_{max}$, so that they give the results corresponding to a state transmission regardless of the sensor's action. We can solve the average reward infinite horizon problem for this MDP using the relative value iteration algorithm, as we satisfy a sufficient condition for convergence, i.e. there exists a state $x \in \mathcal{X}$ that is reachable for every other state in $\mathcal{X}$ under all policies \cite{mahadevan_average_nodate}. As we enforce a transmission at least every $n_{max}$ steps, we can reach a state $(s,s_m,1)$ from every other state, as long as the Markov chain at the source is communicating. If it is not the case, we can simplify the problem by only considering the communicating states. We initialize the value function $v_1^k(x)=0$, $\forall x \in X$. At each step $k$, the algorithm repeatedly applies the following operator to $v_1^k$ until a stopping condition is~satisfied:
\begin{equation} \label{eq:operator1}
    (\Gamma^k V)(x) = \max_{a \in \mathcal{A}^1} R_1^k(x,a) + \mathds{E}_{x' \sim P^a_1}[V(x')] - V(0).
\end{equation}
From the resulting function $v_1^k$, we extract the optimal transmission policy $\pi^k_1$ given the current estimation policy $\pi_2^{k-1}$, which maximizes the value function of player 1 in every~state. 

Afterwards, the transmission policy is fixed and the optimal estimation policy is found. From the point of view of the monitor, we have an MDP denoted by $M_2^k=(\mathcal{Y},\mathcal{A}^2 ,R_2^k ,P_1^k)$. The state space $\mathcal{Y}$ is formed by the set of states $y= (s_m, n) \in \mathcal{S} \times \{ 0,1,...,n_{max}-1 \}$. The action space $\mathcal{A}^2=\mathcal{S}$. The reward function outputs an expectation given the belief over states 
\begin{align} \label{eq:rewards_2}
    R_2^k((s_m,n),a)= 
    \begin{cases}
        b^k(s_m,n)_a-\lambda & n=0 \\
        b^k(s_m,n)_a & n>0 \\
    \end{cases}
\end{align}
where the belief of the monitor, using implicit information is $b^k(s_m,0) = e_{s_m}$ and $b^k(s_m,n) \propto  (P^Tb^k(s_m,n-1) \circ (1-\pi^k_1(\cdot,s_m,n))$, where "$\circ$" represents element-wise multiplication. The transition function is $P_2^k$ and is independent of the action taken, but depends on the transmission policy and is defined as 
\begin{align}
    &P_2^k(\{s_m,n\},\{s_m,n+1\}) = \nonumber\\
    &\qquad \qquad = \Big \langle P^T b^k(s_m,n),1-\pi^1_k(\cdot,s_m,n+1) \Big \rangle \\
    &P_2^k(\{s_m,n\},\{s_m,0\}) \nonumber\\
    &\qquad \qquad= \Big \langle P^T b^k(s_m,n),\pi^1_k(\cdot,s_m,n+1) \Big \rangle.
\end{align}
The independence of the transition probabilities from the action taken confirms that, to maximize this MDP's value function, it is sufficient to maximize the immediate reward at each time step. The relation in (\ref{eq:rewards_2}) suggests that we can do this by setting $\pi_2^k(s_m,n)=\argmax_{a \in \mathcal{A}^2} (b^k(s_m,n))_a$.

\begin{algorithm}[t]
\caption{Alternating Policies Algorithm}\label{alg:alternating}
\begin{algorithmic}[1]
\State $\pi^2_0(s_m,n) \gets (P^n)^Te_{s_m}$ \Comment{initialize monitor policy}
\State $k \gets 0$
\While{$J_k \neq J_{k-1}$} \Comment{loop until convergence}
\State $k \gets k + 1$
\State $\pi^1_k \gets RVI(\Gamma, \pi^2_{k-1})$ \Comment{improve sensor policy}
\State $\pi^2_k(s_m,n) = \argmax_{a \in \mathcal{A}^2} (b^k(s_m,n))_a$ \Comment{improve monitor policy}
\EndWhile
\end{algorithmic}
\end{algorithm}

The next theorem provides a convergence result for the alternating policies algorithm.
\begin{theorem}\emph{
    The infinite horizon average reward of the system $J^k=J(\pi_1^k,\pi_2^k)$ convergences as $k \to \infty$ and the policies obtained upon convergence form a Nash equilibrium.}
\end{theorem}
\begin{IEEEproof}
See Appendix A of supplementary material.
\end{IEEEproof}

\begin{remark}
Note that Problem 1 might possess multiple Nash equilibria, and Theorem 1 guarantees the convergence to one of these equilibria. A question that arises in relation to such a Nash equilibrium is whether it is globally optimal. Answering this question requires development of further techniques such as the one proposed in the next section.
\end{remark}

\section{Occupancy-State Algorithm} \label{sec:continuous-MDP-solution}
In this section, we propose an algorithm based on the notion of occupancy state that finds a globally optimal solution to Problem 1 \cite{dibangoye_optimally_2016}. Let an occupancy-state MDP be denoted by $\tilde{M}=(\tilde{\mathcal{S}},\tilde{\mathcal{A}},\tilde{P}, \tilde{R})$, where $\tilde{\mathcal{S}} \in \Delta(\mathcal{S})$ is the state space such that each state is a belief over states of the Markov chain; $\tilde{\mathcal{A}}: \mathcal{S} \mapsto \{0,1\}$ is the action space, where each action represents a decision rule of the sensor, mapping from the observed state of the sensor to the action taken; such a decision rule is represented by a vector of dimensions $|\mathcal{S}| \times 1$, where the $n$th element of this vector is the action taken by the sensor when the Markov state is $n$; $\tilde{P} \in \tilde{\mathcal{S}} \times \tilde{\mathcal{A}} \times \tilde{\mathcal{S}} \mapsto [0,1]$ is the state transition probability function given by $\tilde{P}(\tilde{s},\tilde{a},\tilde{s}')= \sum_{s \in \mathcal{S}} \tilde{s}_s \mathds{1}(P^Tb'(\tilde{s}, \tilde{a}, s) = \tilde{s}')$ and $b'(\tilde{s}, \tilde{a},s)=\tilde{a}_s e_s + (1-\tilde{a}_s) \frac{(1-\tilde{a}) \circ \tilde{s}}{|(1-\tilde{a}) \circ \tilde{s}|_1}$ is the post-transmission belief given a known decision rule; and finally $\tilde{R} \in \tilde{\mathcal{S}} \times \tilde{\mathcal{A}} \mapsto \mathds{R}$ is the reward function, mapping beliefs and sensor decision rules to rewards, i.e., $\tilde{R}(\tilde s,\tilde a)= \sum_{s\in \mathcal{S}} \tilde{s}_s [\tilde{a}_s(1-\lambda) + (1-\tilde{a}_s) \mathds{1}(\argmax_i(b'(\tilde{s},\tilde{a},s)_i=s)] $, where the $\argmax$ is the optimal monitor action as proved in Section IV. 

We can use this formulation to define the following problem, the solution of which provides us with a solution to the original Problem 1. We define
\begin{align}
&\hspace{-0.6cm}\text{Problem 2:} \nonumber\\
&\underset{\tilde{\pi} \in \tilde{\Pi}}{\maximize} \mathds{E}_{\sim b_0,\tilde{P},\tilde{\pi}} \left [ \lim_{T \to \infty}  \frac{1}{T} \sum_{t=0}^{\infty}  \tilde{R}(\tilde{s}_t, \tilde{\pi}(\tilde{s}_t)) \right ],
\end{align}
where $\tilde{\Pi}$ is the set of Markovian belief-dependent policies for $\tilde{M}$. During each time-step at execution, the sensor obtains the current decision rule by taking $\pi_t=\tilde{\pi}(b_t)$, and then acts according to $a^1_t=(\pi_t)_s$. Algorithm~\ref{alg:occupancy} summarizes the occupancy-state algorithm.

\begin{algorithm}[t]
\caption{Occupancy-State Algorithm}\label{alg:occupancy}
\begin{algorithmic}
\State $\tilde{\pi} \gets RVI\; Q-learning(Problem 2)$ \Comment{solve Problem 2 to obtain the policy}
\State $b \gets b_0$; \Comment{initialize belief using initial state distribution}
\Loop
\State $s \gets sample(P^T e_s)$ \Comment{sensor observes the Markov state}
\State $\pi \gets \tilde{\pi}(b)$ \Comment{obtain the decision rule from the policy}
\If{$\pi(s) = 1$} \Comment{if the sensor transmits}
    \State $y = s$ \Comment{the monitor observes the state}
    \State $b \gets e_y$ \Comment{the monitor updates its belief}
\Else
    \State $y = \epsilon$ \Comment{the monitor does not receive a message}
    \State $b \gets normalize(b \circ (1 - \pi(s)))$ \Comment{bayesian update of the monitor belief}
\EndIf
\State $a^2 = \argmax b$ \Comment{monitor estimates state}
\State $b \gets P^Tb$ \Comment{account for new time-step transition}
\EndLoop
\end{algorithmic}
\end{algorithm}

The next theorem shows that a global optimal solution can be found by the occupancy-state algorithm.
\begin{theorem}\emph{
The transformation of the original two-player problem to the occupancy-state single-player problem is without loss of optimality, and the generated policies are also optimal for Problem 1.}
\end{theorem}
\begin{IEEEproof}
See Appendix B of supplementary material.
\end{IEEEproof}

\begin{remark}
As globally optimal solutions express a stronger solution concept than Nash equilibria, the performance of the solution obtained by Algorithm 2 is better than or equal to that of of the solution obtained by Algorithm 1. Nevertheless, Algorithm 2 can be computationally expensive. The next section focuses on development of a simple algorithm that exploits implicit information and can find a near-optimal solution.
\end{remark}

%We can do the same with our two-player game, as the differences between our game and dec-PODMP do not invalidate the equivalence.

%(What kind of optimality is this supposed to approximate?)

 %It is possible for us to restrict to deterministic policies because the set of optimal policies contains at least one deterministic policy.

%could say something about convexity

%In particular, it knows that the transmission policy is such that a transmission occurs only if the current state is the one with the highest probability.

\section{Heuristic Algorithm}
In this section, we propose a heuristic algorithm that can find a near-optimal solution to Problem 1 with low complexity. In this algorithm, at each time-step, the sensor sends a message to the monitor if and only if the state of the source is not the one with the highest pre-transmission probability in the monitor's belief. Since the sensor has access to all the information that the monitor has access to, this is feasible as it can reason about the monitor's belief. The monitor then guesses correctly because it has received a message containing the current state of the source. If no message is received, the monitor can exploit the implicit information. This eliminates any ambiguity and allows the monitor to always estimate the correct state. Algorithm~\ref{alg:heuristic} summarizes this heuristic method. In this algorithm, $b_0 \in \Delta(\mathcal{S})$ represents the initial distribution over states of the Markov chain, $s$ is the state of the Markov chain, $b$ is the pre-transmission belief, $y$ is the message sent, and $a^2$ is the monitor's action.

\begin{algorithm}[t]
\caption{Heuristic Algorithm}\label{alg:heuristic}
\begin{algorithmic}[1]
\State $b \gets b_0$ \Comment{initialize belief using initial state distribution}
\Loop
\State $s \gets sample(P^T e_s)$ \Comment{sensor observes the Markov state}
\If{$s \neq \argmax b$} \Comment{transmission only occurs if the monitor would guess incorrectly otherwise}
    \State $y = s$ \Comment{the monitor observes the state}
    \State $b \gets e_y$ \Comment{the monitor updates its belief}
\Else
    \State $y = \epsilon$ \Comment{the monitor does not receive a message}
    \State $b \gets e_{\argmax_i b_i}$ \Comment{the monitor updates its belief to the natural vector corresponding to the only state that would not result in transmission}
\EndIf
\State $a^2 = \argmax b$ \Comment{monitor estimates state}
\State $b \gets P^Tb$ \Comment{account for new time-step transition}
\EndLoop
\end{algorithmic}
\end{algorithm}

The next theorem shows that our heuristic algorithm is optimal if perfect reconstruction at the monitor is required.
\begin{theorem}\emph{
    The proposed heuristic algorithm obtains an optimal solution that minimizes the average communication frequency subject to the perfect reconstruction constraint.}
\end{theorem}
\begin{IEEEproof}
See Appendix C of supplementary material.
\end{IEEEproof}

\begin{remark}
Note that in many safety critical applications it is desired to have a perfect reconstruction at the monitor. The proposed heuristic algorithm achieves this optimally by directly exploiting implicit information. Furthermore, it is worth mentioning that our numerical results presented in the next section confirm that the performance of the proposed heuristic algorithm is in fact very close to the performance of the globally optimal solution. 
\end{remark}

% Furthermore, there exists a transmission cost $C \geq 1$ such that this is the optimal policy to solve Problem 1 with transmission cost $c_T \leq C$.

%Behaviour for $0 \leq c < 1$ is transmit for all states but 1. (COULD TRY TO PROVE) Pareto vs what we did with R -c 

\section{Numerical results}
In this section, we present our numerical results. We compare our policies (called \texttt{alternate},\texttt{occupancy} and \texttt{heuristic}) with two other policies adopted in \cite{salimnejad_real-time_2023}, i.e., a uniform policy in which the sensor transmits a message every $u \in \mathds{Z}^+$ time steps (called \texttt{uniform}) and a randomized stationary policy (called \texttt{randomized}), where at each time-step, a transmission occurs with probability $p_{tx}$, independently from the evolution of the system. Note that \cite{salimnejad_real-time_2023} deals with Markov chains with transition matrices in form $P=qI-p(J-I)$, where $I$ is the identity matrix, $J$ is an all-ones matrix and $p,q \in [0,1]$. In their setting, the monitor keeps guessing the last state it received until it receives a new one. We consider general Markov chains without any restrictions, so we modify their policies so that the monitor takes actions $a^2(s_m,n)=\argmax_i ((P^n)^Te_{s_m})_i$. These policies do not exploit implicit information, as the transmission are not based on the state of the source. We also compare our policies with a modified heuristic algorithm (called \texttt{heuristic no implicit}) that neglects the implicit information, and forces the monitor to take the same actions as above: $a^2(s_m,n)= \argmax_i ((P^n)^Te_{s_m})_i$).

\begin{figure}
\begin{tikzpicture}
\begin{axis}[
    xlabel={Average Channel Utilization},
    ylabel={Average Correct Reconstruction},
    xmin=0, xmax=1,
    ymin=0.4, ymax=1,
    xtick={0.0,0.1,0.2,0.3,0.4,0.5,0.6,0.7,0.8,0.9,1.0},
    ytick={0.4,0.5,0.6,0.7,0.8,0.9,1.0},
    legend pos=south east,
    ymajorgrids=true,
    xmajorgrids=true,
    grid style=dashed,
]

\addplot[
    color=green,
    mark=x,
    ]
    table {belief.dat};

\addplot[
    color=purple,
    mark=x,
    ]
    table {implicit.dat};

\addplot[
    color=brown,
    mark=x,
    ]
    table {VI.dat};
    
\addplot[
    color=blue,
    mark=x,
    ]
    table {uniform.dat};

\addplot[
    color=red,
    mark=x,
    ]
    table {whenNotMax.dat};

\addplot[
    color=black,
    mark=x,
    ]
    table {randomized.dat};
    
    \legend{occupancy,heuristic,alternating,uniform,heur. no impl.,randomized}

\end{axis}
\end{tikzpicture}

\caption{Reconstruction performance vs communication cost trade-off for the various schemes considered in the paper. %\texttt{alternate}, \texttt{occupancy} and \texttt{heuristic} represent the policies from Sections IV, V and VI, respectively. \texttt{heur. no impl.} is a different version of \texttt{heuristic} where the monitor does not use the implicit information. \texttt{uniform} and \texttt{randomized} are adapted from \cite{salimnejad_real-time_2023}.
}
\label{fig:reward_vs_communication}
\end{figure}
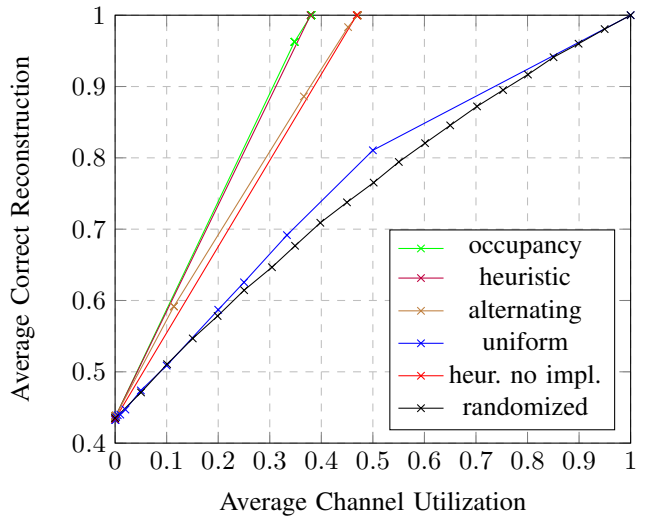

Fig. \ref{fig:reward_vs_communication} shows the trade-off between the average correct reconstruction probability and the average rate of communication. Each line represents a different algorithm and each point a specific solution, obtained with a different value of $\lambda$. \texttt{randomized} performs the worst, having the worst average correct reconstruction probability for all average channel utilization values. \texttt{uniform} is slightly better at medium average channel utilization values. The only way to obtain a point with perfect reconstruction with average channel utilization $<1$ is for the sensor to always transmit when the monitor would have otherwise been wrong, as \texttt{heuristic} and \texttt{heuristic no implicit} do, while \texttt{uniform} and \texttt{randomized} do not reason about the monitor's state to decide whether to transmit. Then, the average channel utilization at perfect reconstruction depends on the quality of the monitor's beliefs, where a more accurate belief reduces the average channel utilization needed to achieve perfect reconstruction. \texttt{heuristic no implicit} achieves a point that is not the most leftwards in the plot as its monitor calculates the beliefs sub-optimally, neglecting the implicit information. \texttt{heuristic} adopts the optimal belief, leading to the most leftwards point. These two algorithms only have one point (policy) each in the graph as they do not have adjustable parameters (in the figure we connect those points to the leftmost zero-transmission point as point connecting on the line can be achieved through time-sharing). \texttt{alternating} matches \texttt{heuristic no implicit} in performance for perfect reconstruction. We can show that such point is a Nash equilibrium of the player's policies, but it is not the globally optimal solution. At lower average channel utilization, this algorithm performs better than \texttt{heuristic no implicit} as the dynamic programming-based policy of the sensor schedules samples optimally, given the estimation policy and the transmission cost parameter $\lambda$. \texttt{occupancy} achieves the best trade-off boundary. \texttt{occupancy} and \texttt{heuristic}'s performance at perfect reconstruction is the same, which is compatible with our theoretical result. \texttt{occupancy} provides slightly better performance at lower channel utilization, which is attributed to a more intelligent scheduling. 

The algorithms can be grouped into 3 classes in terms of performance. The first one includes \texttt{randomized} and \texttt{uniform}. These algorithms transmit information less intelligently, i.e., the sensor does not reason about the behavior of the monitor. Then, \texttt{heuristic no implicit} and \texttt{alternating} transmit information more intelligently, being more likely to transmit when this adds more information to the monitor, though the implicit information is not used or used sub-optimally as \texttt{alternating} is not guaranteed to converge to a global optimum. Lastly, \texttt{heuristic} and \texttt{occupancy} use the implicit information optimally and achieve the best performance. Note that some algorithms are only able to find policies corresponding to very few points on the trade-off curve. This is because the objective function is the average reward minus transmission cost. We are not directly targeting the trade-off between reward and communication in a way that would allow us to find a continuous Pareto optimal boundary. In our formulation, different values of transmission cost can lead to the same policy and so the same point in Fig. 1. However, we can achieve any point on the line joining any two points on the graph via randomized time sharing.

\section{Conclusion}
In this paper, we developed a framework for finding solutions to the problem of remote estimation of Markov processes over costly channels without neglecting implicit information apriori. First, we proposed an algorithm that alternates between optimizing the transmission and estimation policies, and guarantees a Nash equilibrium. We showed that this algorithm performs vastly better than other policies previously explored in the literature. Then, we proposed the occupancy state formulation, which transforms the original two-player problem into a single-player MDP, and guarantees a globally optimal solution. Lastly, we proposed a simple heuristic transmission policy, in which the messages are sent when the monitor would have otherwise guessed incorrectly, which is optimal in terms of minimizing communication costs subject to the perfect reconstruction constraint. 

\bibliographystyle{IEEEtran}
\bibliography{references_manual}

%%
%% where we here have assumed the existence of the files
%% definitions.bib and bibliofile.bib.
%% BibTeX documentation can be obtained at:
%% http://www.ctan.org/tex-archive/biblio/bibtex/contrib/doc/
%%%%%%

%% Or you use manual references (pay attention to consistency and the
%% formatting style!):
%%\begin{thebibliography}{9}

% \bibitem{Laport:LaTeX}
% L.~Lamport,
%   \emph{\LaTeX: A Document Preparation System,} 
%   Addison-Wesley, Reading, Massachusetts, USA, 2nd~ed., 1994. 

% \bibitem{GMS:LaTeXComp}
% F.~Mittelbach, M,~Goossens, J.~Braams, D.~Carlisle, and
% C.~Rowley, \emph{The {\LaTeX} Companion,} Addison-Wesley,
% Reading, Massachusetts, USA, 2nd~ed., 2004.

% \bibitem{oetiker_latex}
% T.~Oetiker, H.~Partl, I.~Hyna, and E.~Schlegl, \emph{The Not So Short
%   Introduction to {\LaTeX2e}}, version 5.06, Jun.~20, 2016. [Online].
%   Available: \url{https://tobi.oetiker.ch/lshort/}

% \bibitem{typesetmoser}
% S.~M. Moser, \emph{How to Typeset Equations in {\LaTeX}}, version 4.6,
%   Sep. 29, 2017. [Online]. Available:
%   \url{http://moser-isi.ethz.ch/manuals.html#eqlatex}

% \bibitem{IEEE:pdfsettings}
% IEEE, \emph{Preparing Conference Content for the IEEE Xplore Digital
%   Library.} [Online]. Available:
%   \url{http://www.ieee.org/conferences_events/conferences/organizers/pubs/preparing_content.html}

% \bibitem{IEEE:AuthorToolbox}
% IEEE, \emph{Author Digital Toolbox.} [Online.] Available:
%   \url{http://www.ieee.org/publications_standards/publications/authors/authors_journals.html}

% \end{thebibliography}

\onecolumn

%\appendix

\section*{Appendix A:\\Proof of Theorem 1} 
\label{Proof of Theorem 1}
\begin{IEEEproof}
Step-5 of Algorithm \ref{alg:alternating} is a maximization over transmission policies of the value function of $M_1^k$ which is given by $
    v^k_1(x,\pi_1)= R^k_1(x,\pi^1(x)) + \sum_{x' \in \mathcal{X}} P_1^k(x,x')V(x')
$,
and we can write the objective function of Problem 1 as a function of the transmission policy, given the estimation policy in Step-5 as $J_1^k(\pi_1) = \lim_{T \mapsto \infty} \frac{1}{T} \mathds{E}_{x_0} [v^1_k(x_0,\pi^1)]$, where $x_0$ is the initial distribution over sensor states. Step-6 maximizes the value function of $M^k_2$, $ 
    v^k_2(y,\pi_2)= R^k_2(y,\pi^2(y)) + \sum_{y' \in \mathcal{Y}} P_2^k(y,y')V(y')
$. We can also write the objective function in Step-6 as a function of the estimation policy as: $J_2^k(\pi_2) = \lim_{T \mapsto \infty} \frac{1}{T} \mathds{E}_{y_0} [v^2_k(y_0,\pi^2)]$, where $y_0$ is the initial distribution over monitor states.

Both of steps 5 and 6 are guaranteed to return the best transmission and estimation policies, respectively given the other player's policy. This is equivalent to maximizing their respective value functions in all states. As we are operating in the infinite horizon undiscounted setting, a policy resulting in an increase in the value of one state must also increase the value of every other state that communicates with it. As a result, when we improve the policies in step 5 and 6  the value function for all reachable states must be greater or equal than before. More formally $J(\pi^1_k,\pi^2_{k-1}) = \max_{\pi_1 \in \Pi_1} J^k_1(\pi^1) \geq J(\pi^1_{k-1},\pi^2_{k-1})$ and $J(\pi^1_k,\pi^2_k) = \max_{\pi_2 \in \Pi_2} J^k_2(\pi^2) \geq J(\pi^1_k,\pi^2_{k-1})$, where $J$ is the objective function as a function of the two policies. The algorithm at every iteration of two policy improvements returns a value of the objective function greater or equal to the previous one, that is it forms a monotonically non-decreasing sequence of $J^k=J(\pi^1_k,\pi^2_k)$. Given that the rewards are bounded, this sequence is bounded above; and hence, our algorithm will converge asymptotically to a finite value. Combined with the finite set of actions in our problem, the convergence will take place after a finite number of steps. Given that there is a $K$ for which $\pi^1_K = \pi^1_{K-1}$, then $\pi^1_h=\pi^1_K$ and $\pi^2_{h-1} = \pi^2_{K-1}$, $\forall h > K$ and given that there is a $K$ for which $\pi^2_K = \pi^2_{K-1}$, then $\pi^1_h=\pi^1_K$ and $\pi^2_h = \pi^2_K$, $\forall h > K$. This is because once one of the two policy improvement steps returns the same policy in two successive iterations, the following sub-step that follows will solve the same sub-problem that it solved in the previous iteration and so the future sub-policies are constant. Lastly, if we do reach this situation, the policies will form a Nash equilibrium, i.e. by fixing one of the policies we cannot achieve a better objective function value by changing the other policy for either of the policies, as otherwise the policies would be changing during steps 5 and 6.
\end{IEEEproof}

\section*{Appendix B:\\Proof of Theorem 2}
\label{Proof of Theorem 2}
\begin{IEEEproof} 
Note that a finite-horizon dec-POMDP can be optimally solved by recasting the problem into a single-player continuous-state MDP~\cite{dibangoye_optimally_2016}. The occupancy state at each time step represents the distribution over the possible underlying states and joint histories given the policy followed so far. The actions are decision rules mapping from the local histories to the local actions of the original game. Our system forms the information state MDP $\hat{M}=(\hat{\mathcal{S}},\hat{\mathcal{A}},\hat{T},\hat{R})$, where information states are more compact representations of the occupancy state that represent the common information available to both agents at each time-step.

The occupancy states at time $t$ are vectors in the form $(|\mathcal{S}| \times |Z^1|^t \times |\mathcal{A}^1|^{t-1} \times |Z^2|^{t-1} \times|\mathcal{A}^2|^t) \times 1$, where each element represents the probability of one realization $Pr(s,z_{0:t}^1,a_{0:t-1}^1,z_{0:t-1}^2,a_{0:t-1}^2|\hat{\pi})$. Having both $s_t$ and $z^1_t$ in the same vector is redundant so we just keep the latter. Note that we do not include the effect of $z^2_t$ in the occupancy state at time $t$, as this is the observation of player 2, which occurs at a later stage in the time-step and is a deterministic function of $z^1_t$ and $a^1_t$. 
    % Thus, including the observation history of player 2 is redundant. $s_t=z^1_t$, making it is redundant to specify the observation of player 1 at time $t$. We showed that the optimal monitor action is a deterministic policy This leads to the more compact (information) $\hat{s_t} \in \hat{S_t}=|\mathcal{S}| \times |Z^1|^{t-1} \times |A^1|^{t-1} \times |A^2|^{t-1}$. 
    We also notice that the actions of player 1 and the observations of player 2 are observed by both players, so their history can be written in deterministic form. The (information) state becomes $\hat{s}_t=(a_{0:t-1}^1,z_{0:t-1}^2,i_t)$, where $i_t(z_{0:t}^1,a_{0:t-1}^2)=Pr(z_{0:t}^1,a_{0:t-1}^2|a_{0:t-1}^1,z_{0:t-1}^2,\hat{\pi}_{0:t-1})$. 
    
The actions are defined as joint decision rules, where each action is a pair of mappings $(\pi_t^1,\pi_t^2)$ from local histories and actions of the original two-player game to the actions' probabilities, with $\pi^1_t \in ((Z^1)^t \times (\mathcal{A}^1)^{t-1} \times \mathcal{A}^1 \mapsto [0,1])$ and $\pi^2_t \in (Z^2)^t \times (\mathcal{A}^2)^{t-1} \times \mathcal{A}^2 \mapsto [0,1])$. 

The policy $\hat{\pi}$ maps information states to decision rules. For now we consider time dependent stochastic policies to be as general as possible.
    
The transition function $\hat{T}(\hat{s}_t,\hat{s}_{t+1},\pi_t) \in [0,1]$ is defined as the probability of transitioning from the current occupancy state to another given the current joint decision rule, where the formulas are given in 
Appendix B.A. Lastly, the reward function is 

\begin{align}
\hat{R}(\hat{s}_t,\pi_t) = \sum_{s_t \in \mathcal{S}} \quad \sum_{z_{0:t-1}^1\in (Z^1)^t , a_{0:t-1}^2 \in (\mathcal{A}^2)^t} &
Pr(z^1_{0:t},a^2_{0:t-1}|\tilde{s}_t) \nonumber\\
&[ \pi^1_t((z_{0:t}^1,a_{0:t-1}^1),0)\pi^2_t((\{z_{0:t-1}^2,\epsilon\}, a_{0:t-1}^2),s_t) \nonumber\\
&+ \pi^1_t((z_{0:t}^1,a_{0:t-1}^1),1)\left(\pi^2_t((\{z_{0:t-1}^2,s_t\}, a_{0:t-1}^2),s_t) - \lambda\right) ] 
\end{align}

Problem 1 can be rewritten using this MDP as
\begin{equation}
%&\hspace{-0.6cm}\text{Problem 2:} \nonumber\\
\underset{\hat{\pi} \in \hat{\Pi}}{\maximize} \mathds{E}_{\sim b_0,\hat{T},\hat{\pi}} \left [ \lim_{T \to \infty}  \frac{1}{T} \sum_{t=0}^{\infty}  \hat{R}(\hat{s}_t, \hat{\pi}_t(\hat{s}_t)) \right ]
\end{equation}
where $\hat{\Pi}$ is the set of information state dependent time dependent policies. We observe that the reward function is only dependent on the components of the information state (other than the current Markov state) through the decision rule, which is a decision variable. The transition of the distribution over underlying states is given by $Pr(z^1_{t+1}|i_{t+1},\pi_t,a^1_{0:t},z^2_{0:t})
\propto  \sum_{z^1_t} P(z^1_t,z^1_{t+1}) Pr(z_t^1|i_t) Pr(z^2_t|z^1_t,a^1_t) \sum_{z^1_{t-1}} \pi_t^1((z_{0:t}^1,a^1_{0:t-1}),a^1_t)$, as shown in Appendix B.B. Thus it is only dependent on the distribution over underlying states in the previous step ($Pr(z_t^1|i_t)$), the common information ($z_{0:t}^2,a_{0:t}^1$) and the last decision rule ($\pi^t$). This means that we can obtain the same value of J* by utilizing just the distribution over underlying Markov states as an information state, which we refer to as the belief. We can then form a MDP where the state at time $t$ is the belief $\tilde{s}_t=Pr(s_t|\hat{s}_t)$, the action is defined as a decision rule mapping from states to sensor actions for player 1 and as simple action for player 2. The following Lemma is used to simplify this further,

\newtheorem{lemma}{Lemma}
\begin{lemma}
Given a transmission policy $\pi^1$, the optimal action of the monitor is the state corresponding to the highest belief, i.e. $a^{2*}_t=\argmax_{a \in \mathcal{A}^2} Pr(a = s_t|z_{0:t}^2,\pi^1,P)$, where $z^2_{0:t}$ is the observation history of the monitor.
\end{lemma}
\begin{IEEEproof}
The proof is given in Appendix B.C.
\end{IEEEproof}

We note that the actions of the monitor do not affect the transitions of the belief state and as of Lemma 1, the optimal monitor action corresponds to the highest post-transmission belief $b'_t \propto b_t \circ (1 - \tilde{a}^1_t)$ where $\tilde{a}^1_t$ is the sensor's decision rule at time $t$.
As the actions of the monitor are fully defined by the state and actions of this MDP, we can model the monitor as part of the environment, simplifying the MDP further and forming $\tilde{M}$.

Lastly, we need to show that we can limit ourselves to deterministic Markovian policies without loss of optimality. The feasibility of finding an optimal solution requires that there exists an optimal solution that is Markovian (stationary and dependent on the MDP's state). This is the case, as in an infinite horizon average reward MDP problem, there always exists a randomized Markovian policy (a stationary policy that's a function of the Markov state) that matches the value of the best history dependent policy \cite{puterman2014markov}. We can then limit ourselves to stationary Markov policies without loss of optimality. Furthermore, $\tilde{M}$ is unichain, i.e. the Markov chain that forms with any deterministic stationary policy contains a single recurrent class and some transient states, assuming the Markov chain at the source is communicating. If this is not the case, once the system enters a recurrent class we can view this as a system with a smaller communicating Markov chain at the source, thus we still have a deterministic policy. As a result we can solve the problem approximately using RVI Q-learning \cite{abounadi_learning_2001}, which we adapted to deal with the continuous state space by using a neural network as in Deep Q-learning \cite{mnih_human-level_2015}. The original tabular representation of RVI Q-learning converges to the global optimum asymptotically in finite-state settings. In reality, for communication rates $>0$, at execution the MDP $\tilde{M}$ visits a finite set of beliefs, however during the learning process this set is not known and we have to rely on generalization to unseen states. 
\end{IEEEproof}

\subsection{Derivation of information state update} \label{app:Derivation of information state update}

The probability of the sensor's action history in step $t+1$ being $a^1_{0:t}$, given that the current information state $\hat{s}_t$ and the current history dependent joint decision rule $\pi_t$ is 
\begin{align*}
&Pr(a^1_{0:t}|\hat{s}_t,\pi_t) = \sum_{z^1_{0:t},a^1_{0:t-1}}Pr(z^1_{0:t},a^1_{0:t-1}|\hat{s}_t)\pi^1_t((z^1_{0:t},a^1_{0:t-1}),a^1_t).
\end{align*}

The probability of the monitor's observation history in step $t+1$ being $z^2_{0:t}$, given that the current information state $\hat{s}_t$ and the current history dependent joint decision rule $\pi_t$ is 

\begin{align*}
&Pr(z^2_{0:t}|\hat{s}_t,\pi_t) = \sum_{z^1_{0:t},z^2_{0:t-1},a^1_{0:t-1}}Pr(z^1_{0:t},z^2_{0:t-1},a^1_{0:t-1}|\hat{s}_t) \pi^1_t((z^1_{0:t},a^1_{0:t-1}),\mathds{1}(z^2_t \neq \epsilon)).
\end{align*}

$i_{t+1}$ can be defined recursively as a deterministic function of $i_t$, $a^1_{0:t}$, $z^2_{0:t}$ and the current joint decision rule $\pi_t$

\begin{align*}
&i_{t+1}(z_{0:t+1}^1,a_{0:t}^2) \\
&\propto Pr(z_{0:t+1}^1,a_{0:t}^2|z_{0:t}^1,a_{0:t}^1,z_{0:t}^2,a_{0:t-1}^2,\pi_t)i_t(z_{0:t}^1,a_{0:t-1}^2) Pr(a^1_t,z^2_t|z_{0:t}^1,z_{0:t-1}^2,a_{0:t-1}^1,a_{0:t-1}^2)\\
&=Pr(z_{t+1}^1|z_{0:t}^1,a_{0:t}^1,z_{0:t}^2,a_{0:t-1}^2,\pi_t) Pr(a_t^2|z_{0:t+1}^1,a_{0:t}^1,z_{0:t}^2,a_{0:t-1}^2,\pi_t) 
i_t(z_{0:t}^1,a_{0:t-1}^2) Pr(a^1_t,z^2_t|z_{0:t}^1,z_{0:t-1}^2,a_{0:t-1}^1,a_{0:t-1}^2)
\\
&=P(z^1_t|z^1_{t+1}) \pi^2_t((z^2_{0:t}, a^2_{0:{t-1}}), a_t^2)i_t(z_{0:t}^1,a_{0:t-1}^2) 
 \pi^1_t((z^1_{0:t},a^1_{0:t-1}),a^1_t)[a^1_t\mathds{1}(z^1_t=z^2_t)+(1-a^1_t)\mathds{1}(z^2_t=\epsilon)],
\end{align*}

where the first line is a simplified application of Bayes rule. The combination of these formulas fully specifies the state transitions of $\hat{M}$.

\subsection{Derivation of belief transition}\label{app:Derivation of belief transition}
The following equations show that at time step $t+1$ the belief $Pr(z^1_{t+1}|i_{t+1},\pi_t,a^1_{0:t},z^2_{0:t})$ can be written as a function of the previous belief $Pr(z_t^1|i_t)$ and it's only dependent on $z_{0:t-1}^2,a_{0:t-1}^1$ through the decision rule
\begin{align*}
&Pr(z^1_{t+1}|i_{t+1},\pi_t,a^1_{0:t},z^2_{0:t}) 
    \propto \sum_{z^1_{0:t},a^2_{0:t-1}} Pr(z_{t+1}^1|z_{0:t}^1,a_{0:t}^1,z_{0:t}^2,a_{0:t-1}^2,\pi_t)  i_t(z_{0:t}^1,a_{0:t-1}^2) Pr(a^1_t,z^2_t|z_{0:t}^1,z_{0:t-1}^2,a_{0:t-1}^1,a_{0:t-1}^2,\pi_t) 
    \\
&= \sum_{z^1_{0:t}} P(z^1_t,z^1_{t+1})
    Pr(a^1_t,z^2_t|z_{0:t}^1,z_{0:t-1}^2,a_{0:t-1}^1,\pi_t) \sum_{a^2_{0:t-1}}
        i_t(z_{0:t}^1,a_{0:t-1}^2)
    \\ 
&= \sum_{z^1_{0:t}} P(z^1_t,z^1_{t+1}) Pr(a^1_t,z^2_t|z_{0:t}^1,z_{0:t-1}^2,a_{0:t-1}^1,\pi_t)
    Pr(z_{0:t}^1|i_t)
    \\
&= \sum_{z^1_{0:t}} P(z^1_t,z^1_{t+1}) Pr(z_t^1|i_t) \pi_t^1((z_{0:t}^1,a^1_{0:t-1}),a^1_t)Pr(z^2_t|z^1_t,a^1_t)
    \\
&= \sum_{z^1_t} P(z^1_t,z^1_{t+1}) Pr(z_t^1|i_t) Pr(z^2_t|z^1_t,a^1_t) \sum_{z^1_{t-1}} \pi_t^1((z_{0:t}^1,a^1_{0:t-1}),a^1_t).
\end{align*}

\subsection{Derivation of monitor's best actions}
\label{Proof of the best monitor action}
By construction, the Markov source's transitions are unaffected by the behavior of the two players, $Pr(s_{t+1}=s'|s_t=s,a^1_t=a^1,a^2_t=a^2)=Pr(s_{t+1}=s'|s_t=s)$. The sensor does not observe the actions of the monitor, so these can't affect its future behavior. As the monitor receives no feedback on its actions, these can't affect its future behavior either. The monitor actions thus only affect the immediate rewards of the system. Thus the objective of Problem 1 can be maximized by maximizing immediate rewards at each time step. The monitor action does not affect $c_t$ thus it must maximize $r_t=\mathds{1}(s_t = a^2_t)$, giving that the optimal action is the state with maximum monitor belief at time $t$ following the transmission slot.

\section*{Appendix C:\\Proof of Theorem 3}
\label{Proof of Theorem 3}
\begin{IEEEproof}
We want to find the joint policy with perfect reconstruction, that minimizes the average number of communication packages sent over the long term. This can be formulated as

\begin{align}
&\hspace{-2.2cm}\text{Problem 3:} \nonumber\\
&\underset{\pi \in \Pi}{\minimize} \mathds{E}_{\sim b_0, P, \pi}[c_t|\pi] \nonumber\\
&\subjectto \mathds{E}_{\sim b_0, P, \pi}[r_t|\pi]=1 
\end{align}
where $\Pi$ is the set of joint history-dependent stochastic
policies for players 1 and 2, for optimality. In section 5, we showed that the estimation policy does not affect the communication costs as the monitor actions occur after the transmission slot at each time-step and they do not affect the evolution of the system. Problem 3 can be thus be simplified as finding the transmission policy that minimizes communication costs among those policies allowing perfect reconstruction when combined with some estimation policy. We also showed that the estimation policy that estimates the state with the highest belief maximizes the rewards given a fixed transmission policy. If it is possible to achieve average perfect reconstruction with a given transmission policy, it must be possible to do so with the maximum belief based estimation policy, as it maximizes the rewards, which proves that the maximum belief estimation policy is optimal for Problem 3. 

We now need to find the transmission policy that minimizes costs while achieving perfect reconstruction when combined with the $\argmax b$ estimation policy. The perfect reconstruction constraint requires that the immediate reward in every sensor and monitor state with visitation probability $>1$ be equal to $1$. A sufficient condition is then to set the policy so that the reward is deterministically $1$ in every state of the system, as the policy in states that are never visited does not have any effect on the performance. To achieve this, the sensor must send a message whenever the absence of a message would mean that the monitor would estimate the wrong state. As the policy of the monitor is $\pi^2(b') = \argmax_i b'_i$, where $b'$ is the post-transmission belief, the sensor must transmit for all $s\;s.t.\;s\neq\argmax_i b_i$, where $x$ is the sensor state, which includes the current state $s$. This means that we can only not transmit in one source state, given a fixed belief, as if we did otherwise, the monitor would not know with certainty in which one of the passive states we are. Let the passive set of a belief be the set of states for which no transmission occurs in that belief. Knowing that we have a passive set $\mathcal{X}_p$ of size 0 or 1, we can then write the optimal Bellman equation where we use the results from section 5 and we let the state be the belief:

We now need to find the transmission policy that minimizes costs while achieving perfect reconstruction when combined with the $\argmax b$ estimation policy. The perfect reconstruction constraint requires that the immediate reward in every sensor and monitor state with visitation probability $>1$ be equal to $1$. A sufficient condition is then to set the policy so that the reward is deterministically $1$ in every state of the system, as the policy in states that are never visited does not have any effect on the performance. To achieve this, the sensor must send a message whenever the absence of a message would mean that the monitor would estimate the wrong state. As the monitor must always guess correctly even if it receives no message, there must be at most one state for each belief, for which no message is sent, to avoid ambiguity. If this were not the case, the receiver would not have certainty of the current state when no message is received. Additionally, there is no advantage in transmitting for all Markov states, given a fixed belief, as it conveys no extra information but incurs extra cost. 

We proved that the monitor policy must be the $\argmax$ policy and that for a certain belief, we must have one source state in which the sensor does not transmit, while it transmits for all the others. Thus the sensor transmission policy is based on beliefs, agreeing with our findings from Section 5 and we can find the optimal state in which not to send messages for a given belief using the optimal Bellman equation:

\begin{equation}
    V(b) = \max_{s_a \in \mathcal{S}} b_{s_a}(1+V(P^T e_{s_a})) + \sum_{s \in \mathcal{S} \neq s_a} b_s(1-\lambda+V(P^T e_s))
    = 
    \max_{s_a \in \mathcal{S}} b_{s_a} + \sum_{s \in \mathcal{S} \neq s_a} b_s(1-\lambda) +\sum_{s \in \mathcal{S}}b_sV(P^T e_s)
\end{equation}

where $s_a$ is the state where no transmission occurs. This choice does not affect the future time-steps and it is sufficient to minimize immediate costs. It is clear that $s_a$ should be the state corresponding to the highest belief value, which proves that our heuristic algorithm gives the optimal transmission and estimation policies for perfect reconstruction.
\end{IEEEproof}

\end{document}